\documentclass[12pt]{iopart}
\usepackage[T1]{fontenc}
\usepackage{times}
\usepackage{graphicx}
\usepackage{bm}
\usepackage[colorlinks,
breaklinks,
citecolor=blue,
linkcolor=blue,
]{hyperref}
\usepackage{iopams}
\begin{document}

\title{Nonlinear optomechanics with gain and loss: amplifying higher-order sideband and group delay}

\author{Y. Jiao$^{1,\ast}$, H. L\"u$^{2,3,\ast}$, J. Qian$^{2,\dagger}$, Y. Li$^{4}$, and H. Jing$^{5,\dagger}$}
\address{
$^1$Department of Physics, Henan Normal University, Xinxiang 453007, China \\
$^{2}$Key Laboratory for Quantum Optics, Shanghai Institute of Optics and Fine Mechanics, Chinese Academy of Sciences, Shanghai 201800, China \\
$^{3}$University of Chinese Academy of Sciences, Beijing 100049, China \\
$^{4}$Beijing Computational Science Research Center, Beijing 100084, China \\
$^5$Key Laboratory of Low-Dimensional Quantum Structures
and Quantum Control of Ministry of Education, Department of Physics
and Synergetic Innovation Center for Quantum Effects and
Applications, Hunan Normal University, Changsha 410081, China\\
$^{\ast}$These authors contributed equally to this work. \\
$^{\dagger}$Authors to whom any correspondence should be addressed.}
\ead{jqian@siom.ac.cn, jinghui73@foxmail.com}
\vspace{10pt}

\begin{abstract}
We study the nonlinear optomechanically-induced transparency (OMIT) with gain and loss. We find that (i) for a single active cavity, significant enhancement can be achieved for the higher-order sidebands, including the transmission rate and the group delay; (ii) for active-passive-coupled cavities, hundreds of microsecond of optical delay or advance are attainable for the nonlinear sideband pulses in the parity-time-symmetric regime. The active higher-order OMIT effects, as firstly revealed here, open up the way to make a low-power optomechaical amplifier, which can amplify both the strength and group delay of not only the probe light but also its higher-order sidebands.
\end{abstract}

\pacs{42.50.Wk, 03.65.Ta}

%
%
%
\maketitle
%
%

\section{Introduction}
Cavity optomechanics (COM), exploring radiation-pressure-induced coherent photon-phonon interactions, have led to diverse applications in recent years
\cite{Aspelmeyer2012Quantum,Metcalfe2014Applications,Aspelmeyer2014rmp}, such as weak-force measurement, phonon cooling \cite{Teufel2011Sideband,Chan2011Laser} or lasing \cite{phononlaser1,phononlaser2}, and phonon squeezing \cite{Wollman2015Quantum}. A particularly relevant example to our work here is the optomechanically-induced transparency (OMIT) \cite{Weis2010Optomechanically,Safavi2011Electromagnetically}, which provides a new way to store light in solid-state devices \cite{Zhou2013Slowing,Chang2011Slowing,Fiore2011Storing,Fan2015Cascaded}. In addition, OMIT can be used to study a variety of exotic effects due to the nonlinear nature of COM interactions \cite{Ludwig2012Enhanced,Lemonde2013Nonlinear,Nunnenkamp2013Signatures,Liu2013Parametric,Kronwald2013Optomechanically}. A notable example is the nonlinear higher-order sideband effects in the passive OMIT \cite{Xiong2012Higher,Ma2015Optomechanically,Suzuki2015Nonlinear,Ma2014Tunable}, the signal of which, however, is much weaker in comparison with the probe field. The significant enhancement of nonlinear OMIT effect, including the higher-order sideband signal and the associated higher-order slow light, is useful for more flexible COM control of light.

In parallel with these advances, parity-time(PT)-symmetric phase transitions \cite{Bender1998Real,Bender2013Twofold} have also been observed recently in coupled optical microresonators \cite{Peng2014Parity}, opening up novel applications such as low-power optical diodes \cite{Peng2014Parity,Chang2014Loss}, single-mode lasing \cite{Feng2014Single,Hodaei2014Parity}, and other unconventional optical effects \cite{Lijiahui2016,Peng2014Loss,CPA}. Combining these two fields leads to new PT-assisted hybrid COM devices, revealing unique effects such as PT-symmetric phonon lasing \cite{phononlaser2} and PT-broken chaos \cite{LvXY2015}. An inverted OMIT spectrum, i.e. a non-amplifying dip in the otherwise strongly amplifying region, was also identified in the linear process \cite{Jing2015Optomechanically,IEIT}.

In this work, we proceed to study the nonlinear OMIT process in a single active cavity or active-passive-coupled resonators. We show that (i) the second-order sideband exhibits an inverted-OMIT profile in both cases, in particular, the transmission of the second-order sideband pulse can be significantly enhanced in the vicinity of gain-loss balance; (ii) the optical delay or advance of the higher-order sideband can be well adjusted within several hundreds of microsecond, indicating the possibility to engineer the relative group delay of the probe and the higher-order sidebands simultaneously. We note that the active nonlinear OMIT effect, as we studied here, is different from the fast light phenomena by virtue of gain doublet effects \cite{Wang2000fastlight} in atomic systems, in which only the signal pulse itself is generally controlled. As far as we know, the gain-enhanced group delay or advance of nonlinear second-order sideband has not been investigated in previous literatures.

This paper is organized as follows. In Sec. \ref{sec:model}, we present the nonlinear OMIT models for both a single active cavity and an active-passive compound system, with which to calculate the optical spectral responses and the associated group delays of both the probe field and the second-order sideband. Then in Sec. \ref{sec:result}, we analyze in details the results of optical transmissions rates and group delays in a single cavity or an active-passive compound system, respectively, focusing on the roles of the optical gain or the gain-to-loss ratio. Finally in Sec. \ref{sec:conclusion}, we conclude with a brief discussion of the results presented in this work and an outlook of further works.

\section{Theoretical Model} \label{sec:model}

\begin{figure}[tbp]
\centering
\includegraphics[width=0.9\textwidth]{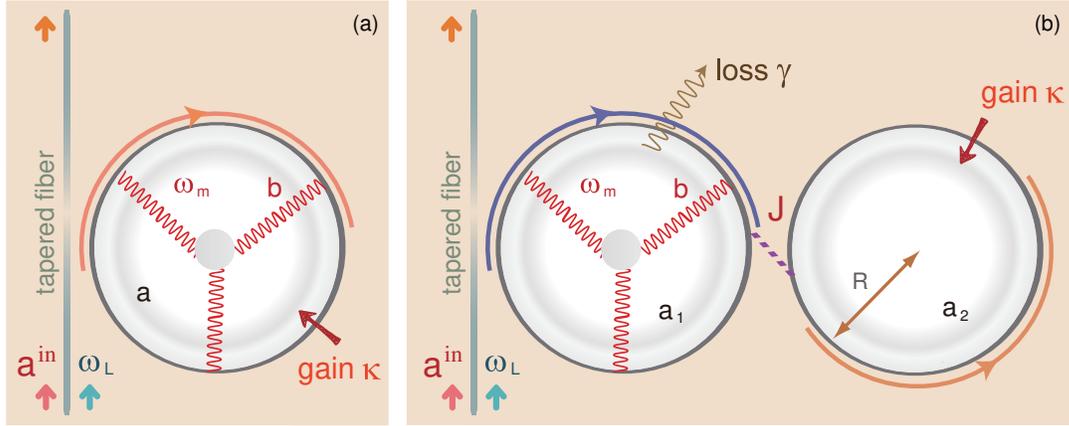}
\caption{(Color online) Schematic of COM in a single active cavity (a) or an active-passive compound structure (b). In (b), the left cavity with an optical decay rate $\gamma$ is coupled with a tapped fiber, and the right cavity with an optical gain $\kappa$ \cite{Peng2014Parity,Jing2015Optomechanically} is coupled to the left cavity with an optical tunneling rate $J$. In both cases, only the cavity coupled to the fiber supports a mechanical mode at frequency $\omega_m$ and is driven by a pump field at frequency $\omega_l$.}
\label{fig1}
\end{figure}

As shown in Fig. \ref{fig1}, for possible comparisons, we consider two different setups consisting of a single active optomechanical cavity \cite{Geli} and an active-passive compound structure, respectively. In both cases, the cavity-mode resonance frequency is $\omega_c$, the cavity (coupled with the tapped fiber) contains a mechanical mode (with the effective mass $m$ and the damping rate $\Gamma_m$) \cite{phononlaser1}, and the COM coupling strength is denoted by $g$. As in Ref. \cite{Peng2014Parity}, the active resonator, fabricated from $\mathrm{Er}^{3+}$-doped silica, can emit photons in the $1550\,\mathrm{nm}$ band, when driven by a laser in the $980\,\mathrm{nm}$ or $1450\,\mathrm{nm}$ bands. In Fig. \ref{fig1}(a), we first focus on the gain-assisted OMIT in a single cavity, including the second-order transmission rate and the associated group delay. In Fig. \ref{fig1}(b), we turn to study the enhancement of the second-order sideband in the vicinity of gain-loss balance in a compound structure. Note that only the photons at the emission band of $1550\,\mathrm{nm}$ can tunnel through the air gap between the resonators, which provides an optical gain $\kappa$ to compensate the optical loss $\gamma$ in the passive resonator \cite{Peng2014Parity}. To observe the OMIT phenomenon, the optomechanical cavity is driven by both a strong pump laser at frequency $\omega_l$ and a weak probe light at frequency $\omega_p$, with the amplitudes $\varepsilon_l = \sqrt{\frac{2\gamma P_l}{\hbar \omega_l}}, ~~~\varepsilon_p = \sqrt{\frac{2\gamma P_{in}}{\hbar \omega_p}}$, respectively, where $P_l$ and $P_{in}$ denote the powers of the control and probe lasers. We are interested in the role of nonlinear COM interactions on the optical responses of the systems, particularly the relative group delay or advance of the probe light and the higher-order sideband pulse. Passive COM setups with single and compound cavities have been extensively investigated in theory and experiments in the past years. However, the study of optomechanical effects with active optomechanics are still in early stages \cite{Geli}, and the effects of gain-assisted optomechanical nonlinearity on high-order OMIT have been little studied.

\subsection{Nonlinear OMIT in an active cavity}

As illustrated in Fig. \ref{fig1}(a), we study the transmission rates of the probe field and the second-order sideband in a single active optomechanical cavity \cite{Geli}. The Hamiltonian of the system can be written at the simplest level as
\begin{equation}
\label{eq:1c}
{H}_1 = \hbar \Delta_l {a}^{\dag} {a} + \frac{{{p}^2}}{2 m} + \frac{1}{2} m \omega_m^2 {x}^2 - \hbar g {a}^{\dag} {a} x + i \hbar \left( \varepsilon_l {a}^\dag + \varepsilon_{p} {a}^{\dag} e^{-i\xi t} - h.c. \right),
\end{equation}
with the optical operator $a$ and the mechanical position or momentum operator $x$ or $p$, and the cavity-pump detuning and the probe-pump detuning, respectively, $\Delta_l \equiv \omega_c - \omega_l,~~ \Delta_p \equiv \omega_p - \omega_c,~~ \xi \equiv \omega_p - \omega_l$. The Heisenberg equations of motion of this system are then
\begin{eqnarray} \label{eq:1c-motion}
\ddot{x} + \Gamma_m \dot{x} + \omega_m^2 x - \frac{\hbar g}{m} a^{\dag} a&=0, \nonumber \\
\dot{a}+(i \Delta_l - igx  - \kappa) a-\varepsilon_l&= \varepsilon_p e^{-i\xi t}.
\end{eqnarray}
The optical or mechanical gain and damping terms are added phenomenologically into the Eq. (\ref{eq:1c-motion}) \cite{phononlaser2, LvXY2015}. Setting all the derivatives as zero leads to the steady-state values
\begin{equation} \label{eq:1c-steady}
x_s = \frac{\hbar g}{m \omega_m^2} {\left| {a_s} \right|^2},  ~~~
a_s = \frac{\varepsilon_l}{-\kappa + i{\Delta _l} - i g x_s}.
\end{equation}
All the operators are then divided into their stationary values and the remaining fluctuations, i.e. $x = x_s + \delta x, a = a_s + \delta a$, and the equations satisfied by the fluctuation terms are
\begin{eqnarray} \label{eq:1c-fluc}
\delta \ddot{x} + \Gamma_m \delta \dot{x} + \omega_m^2 \delta x -\frac{{\hbar g}}{m}(a_{s} \delta a^{\dag} + a_s^{\ast} \delta a + \delta a^{\dag} \delta a)&=0,
\nonumber \\
\delta \dot{a} + (i\Delta_l-igx_s-\kappa)\delta a - ig(a_{s}\delta x+\delta x \delta a)- \varepsilon_p e^{-i\xi t}&=0.
\end{eqnarray}
Here the second-order nonlinear terms such as $\delta x \delta a$ and $\delta a^{\dag} \delta a$ are preserved in Eq. (\ref{eq:1c-fluc}). Now we solve these equations by using the ansatz
\begin{eqnarray} \label{eq:1c-expansion}
\delta a &=& A_{+}^{(1)}e^{-i \xi t} + A_{-}^{(1)} e^{i\xi t}  + A_{+}^{(2)} e^{-2i\xi t} + A_{-}^{(2)} e^{2i\xi t},  \nonumber \\
\delta x &=& X^{(1)} e^{-i \xi t}    + X^{(1) \ast} e^{i\xi t} + X^{(2)} e^{-2i\xi t} + X^{(2)\ast} e^{2i\xi t}.
\end{eqnarray}
Substituting Eq. (\ref{eq:1c-expansion}) into Eq. (\ref{eq:1c-fluc}), and first neglecting the second-order parts in Eq. (\ref{eq:1c-expansion}), leads to the linear response of the system
\begin{eqnarray} \label{eq:1c-linear}
A_{+}^{(1)} &=& \frac{{[\lambda(\xi)\lambda_{s2}(\xi)+i\hbar g^2{\left| {a_{s}} 	 \right|^2}]}\varepsilon_p}{{\lambda(\xi)\lambda_{s1}(\xi)\lambda_{s2}(\xi)
		+i[\lambda_{s1}(\xi)-\lambda_{s2}(\xi)]\hbar g^2{\left| {a_{s}} \right|^2}}}, \\
X^{(1)} &=& \frac{{\lambda_{s2}(\xi)\hbar g a_{s}^{\ast}}\varepsilon_p}{{\lambda(\xi)\lambda_{s1}(\xi)\lambda_{s2}(\xi)+i[\lambda_{s1}(\xi)-\lambda_{s2}(\xi)]\hbar g^2{\left| {a_{s}} \right|^2}}}, \nonumber
\end{eqnarray}
where
$
\lambda(\xi) = m(\omega_m^2-\xi^2-i\xi\Gamma_m),~\lambda_{s1,s2}(\xi) =-i\xi-\kappa \pm (i\Delta_l-igx_s).$
Combining Eqs. (\ref{eq:1c-fluc})-(\ref{eq:1c-linear}) gives the second-order sideband amplitude
\begin{equation}\label{eq:A2}
A_{+}^{(2)} = \frac{{C_1 X^{(1)2} + C_2 A_{+}^{(1)} X^{(1)}}}{{\lambda_{s2}(\xi) [ \lambda(2\xi) \lambda_{s1}(2\xi) \lambda_{s2}(2\xi) + i (\lambda_{s1}(2\xi)-\lambda_{s2}(2\xi))\hbar g^2 {\left| {{a_{s}}} \right|^2}]}}，
\end{equation}
where
$
C_1 = -i\hbar g^4 a_{s} {\left| {{a_{s}}} \right|^2}, ~
C_2 = \hbar g^3 (\lambda_{s2}(2\xi)-\lambda_{s2}(\xi)) {\left| {a_{s}} \right|^2} + i g \lambda_{s2}(\xi) \lambda(2\xi) \lambda_{s2}(2\xi).
$
We note that $A_{+}^{(2)}$, being proportional to $\varepsilon^2_p$, is indeed smaller than the first-order term $A_{+}^{(1)}$ for a weak probe light. The first term or the second term of $A_{+}^{(2)}$ denotes an upconverted-like process of the pump light or the first-order sideband, respectively \cite{Xiong2012Higher}.

\subsection{Nonlinear OMIT in active-passive-coupled resonators}

By including two optical modes $a_{1,2}$ and the mechanical mode containing now in the passive resonator [see Fig. \ref{fig1}(b)], we write the Hamiltonian of this system as
\begin{equation} \label{eq:2c}
{H}_2 = {H}_1+\hbar \Delta_{l} {a}_{2}^{\dag} {a}_{2} -\hbar J ({a}_{1}^{\dag} {a}_{2} + {a}_{2}^{\dag} {a}_{1}),
\end{equation}
with the replacement $a\rightarrow a_1$ for the expression of ${H}_1$. The Heisenberg equations of this compound system are
\begin{eqnarray} \label{eq:2c-motion}
\ddot x &=- {\Gamma _m}\dot x - \omega_m^2 x + \frac{{\hbar g}}{m}a_{1}^{\dag}a_{1}, \nonumber \\
\dot a_{1} &= (-i\Delta_l + igx - \gamma) a_{1} +i J a_{2} + \varepsilon_l + \varepsilon_p e^{-i\xi t}, \nonumber \\
\dot a_{2} &= (-i\Delta_l + \kappa) a_{2} + i J a_{1}.
\end{eqnarray}
Expanding the operators as follows
\begin{eqnarray} \label{eq:2c-expansion}
a_{1} &=& a_{1,s} + A_{1+}^{(1)} e^{-i\xi t} + A_{1-}^{(1)}e^{i\xi t}+A_{1+}^{(2)}e^{-2i\xi t}+A_{1-}^{(2)} e^{2i\xi t}, \nonumber\\
a_{2} &=& a_{2,s} + A_{2+}^{(1)} e^{-i\xi t} + A_{2-}^{(1)}e^{i\xi t}+A_{2+}^{(2)}e^{-2i\xi t}+A_{2-}^{(2)} e^{2i\xi t}, \\
x   &=& x_{s}   + X^{(1)} e^{-i\xi t}  + X^{(1)\ast} e^{i\xi t} + X^{(2)} e^{-2i\xi t} + X^{(2)\ast} e^{2i\xi t}, \nonumber
\end{eqnarray}
where $a_{i,s}$ ($i=1,2$) and $x_{s}$ denotes the steady values, leads to the amplitude of the first-order sideband $A_{1+}^{(1)}$ and $X^{(1)}$, as given in Ref.\cite{Jing2015Optomechanically}, and also the second-order sideband amplitude $A_{1+}^{(2)}$. We find that it has exactly the same form as the single-cavity result, i.e. the Eq. (\ref{eq:A2}), but with the replacements: $a \rightarrow a_1,~~A_{+}^{(1)} \rightarrow A_{1+}^{(1)},$ and
and $$\lambda_{s1} \rightarrow \lambda_{d1}, ~~\lambda_{s2} \rightarrow \lambda_{d2},~~\lambda_{d1,\,d2}(\xi)\equiv- i\xi+\gamma\pm\left( i\Delta_l  -i g x_s+\frac{{J^2}}{{i\Delta_l-i\xi-\kappa}}\right).$$

The output-field expectation value can be obtained by using the standard input-output relation, i.e. $a_1^{out} = a_1^{in} - \sqrt{\gamma} a_1$, where $a_1^{in}$ and $a_1^{out}$ are the input and output field operators, respectively. The transmitted rate of the probe \cite{Jing2015Optomechanically} and the efficiency of the sideband (i.e. the ratio of the amplitude of the output
second-order sideband to the amplitude of the input field) \cite{Xiong2012Higher} then read $$|t_p|^2 =\left|\frac{a_{1}^{out}}{a_{1}^{in}}\right|^2=\left|1-\frac{\gamma}{\varepsilon_p}A_{1+}^{(1)}\right|^2,~~~
\eta=\left|\frac{\gamma}{\varepsilon_p}A_{1+}^{(2)}\right|.$$
Also, taking into account of the second-order sideband frequency $2 \xi + \omega_l$, the group delay of the probe light or the second-order sideband is given by
$$\tau_{g}=\frac{d \mathrm{arg}(t_{p})}{d\xi}|_{\xi=\omega_m},~~~\tau_{g}'=\frac{d \mathrm{arg}(A^{(2)}_{1+})}{d \omega}|_{\xi = \omega_m} =
\frac{d \mathrm{arg}(A^{(2)}_{1+})}{2d\xi}|_{\xi=\omega_m},$$
respectively. In the following simulations, experimentally accessible parameters are chosen, i.e. $\varepsilon_{p}/\varepsilon_{l}=0.05$, (the resonator radius) $R=\omega_c/g= 34.5\mu$m, $m$ = 50 ng, $\omega_c = 1.93 \times 10^5$ GHz, $\omega_m = 2 \pi \times 23.4$ MHz, $\gamma = 6.43$ MHz, $\Gamma_m = 2.4 \times 10^5$ Hz, and $\Delta_l=\omega_c-\omega_l=\omega_m$ or $\Delta_p\equiv \omega_p-\omega_c$. These parameter values of coupled resonators have also been used in such a wide range of fields from nonreciprocal light propagation \cite{Peng2014Parity}, loss-induced revival of lasing \cite{Peng2014Loss}, phonon lasing \cite{phononlaser1,phononlaser2} and phonon diode \cite{Zhang} to low-power chaos \cite{LvXY2015} and ultra-sensitive force measurement \cite{Liu}.

\section{Results and discussions} \label{sec:result}

\subsection{Single-cavity case}

\begin{figure}[tbp]
\includegraphics[width=0.49\textwidth]{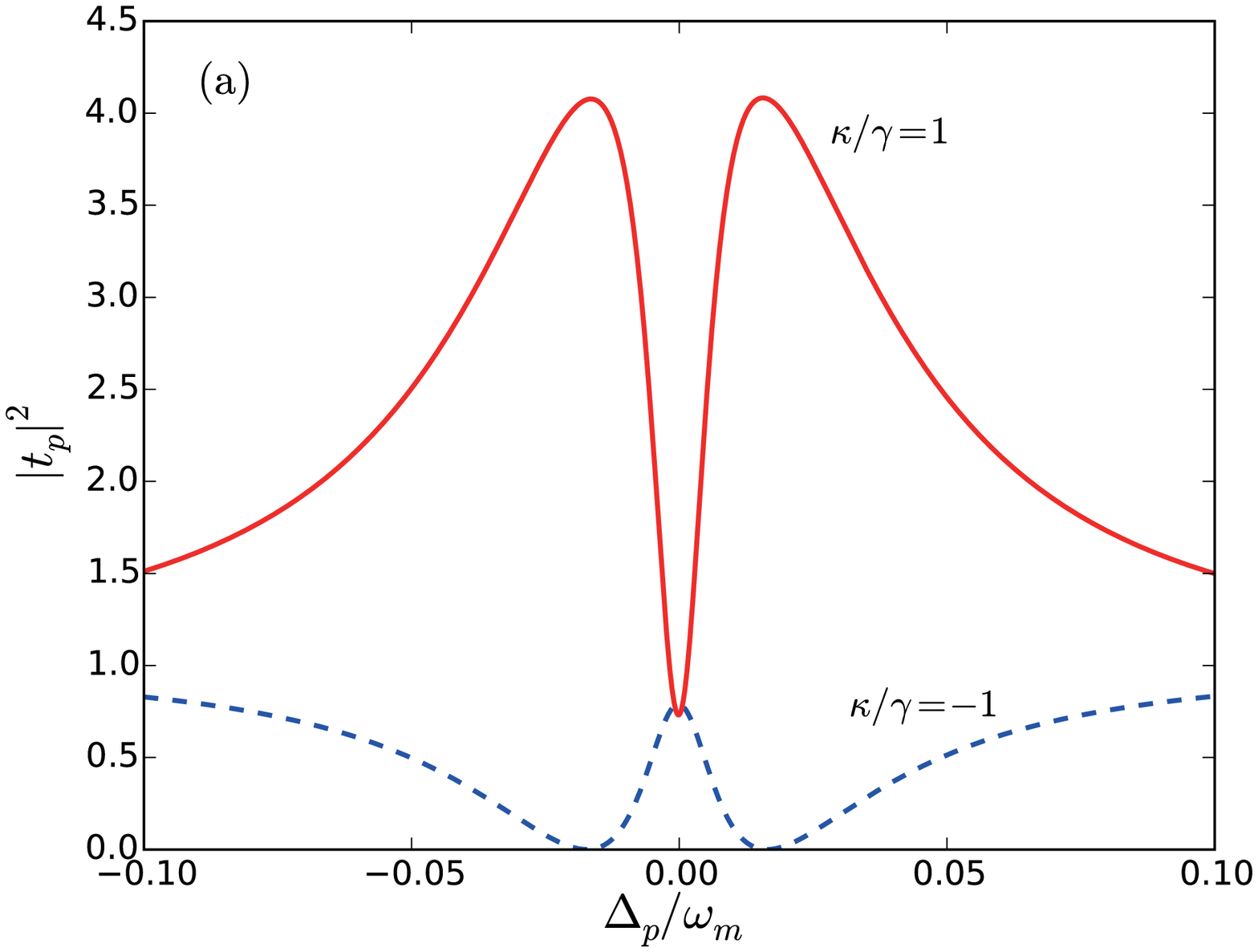}
\includegraphics[width=0.49\textwidth]{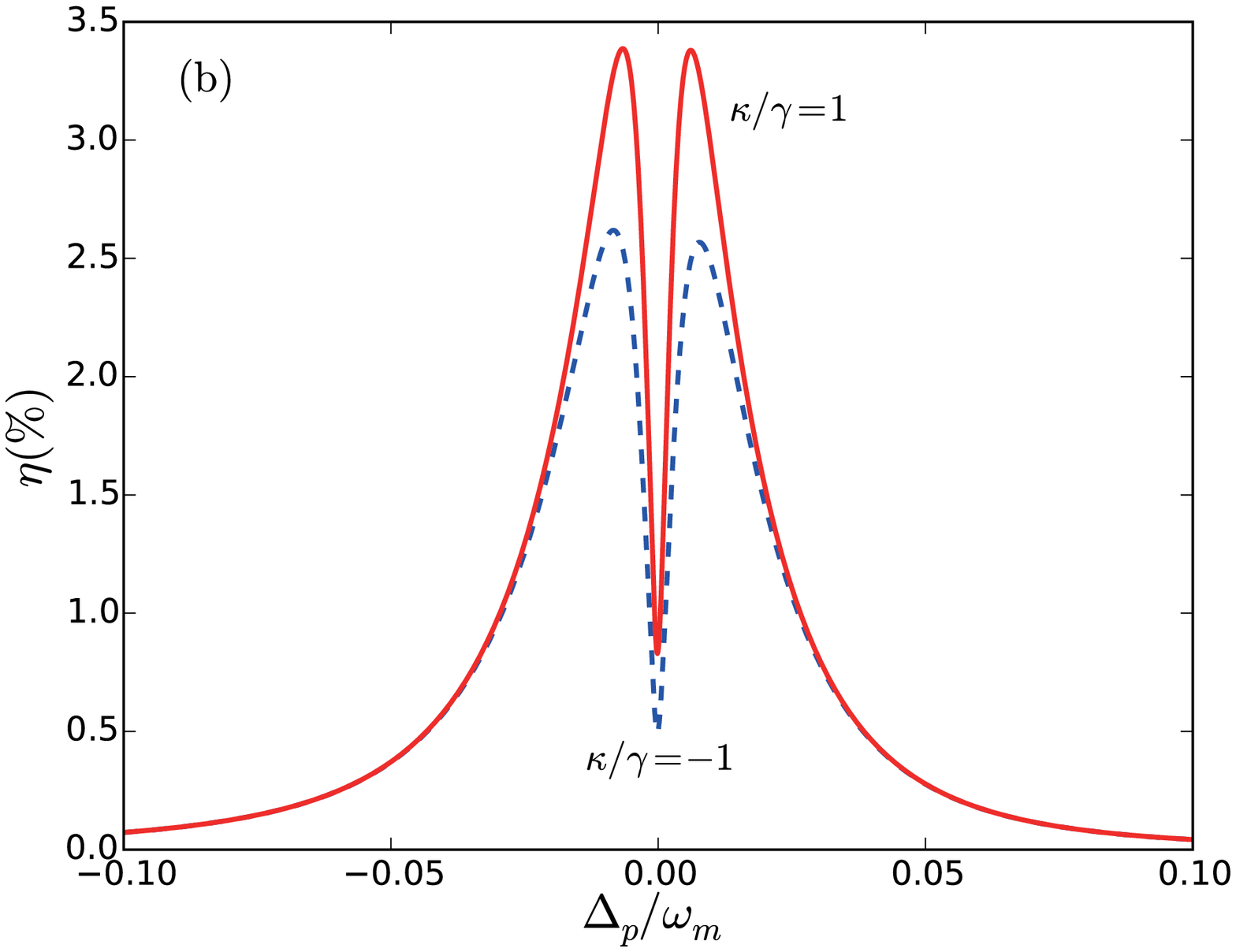} \\
\includegraphics[width=0.49\textwidth]{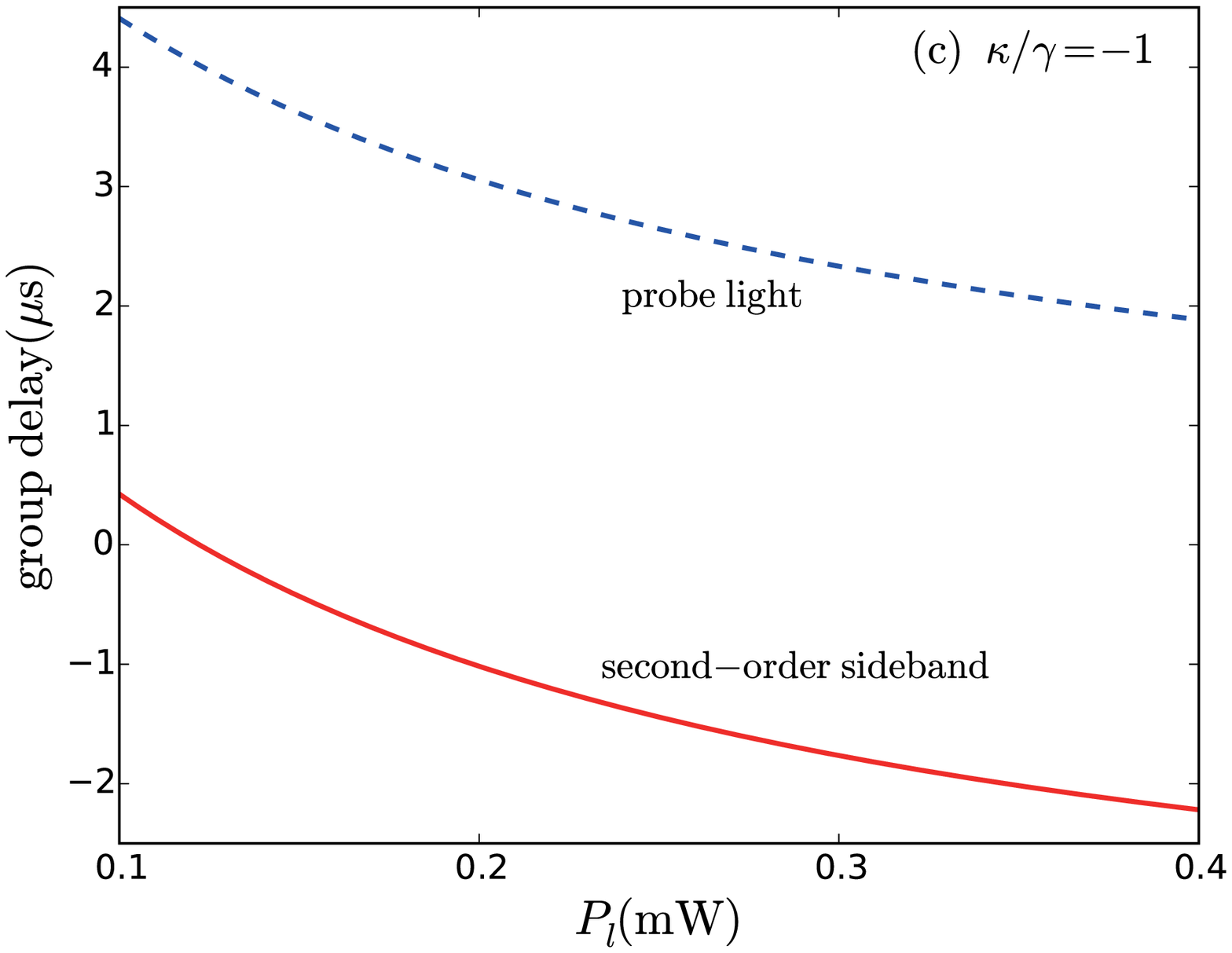}
\includegraphics[width=0.49\textwidth]{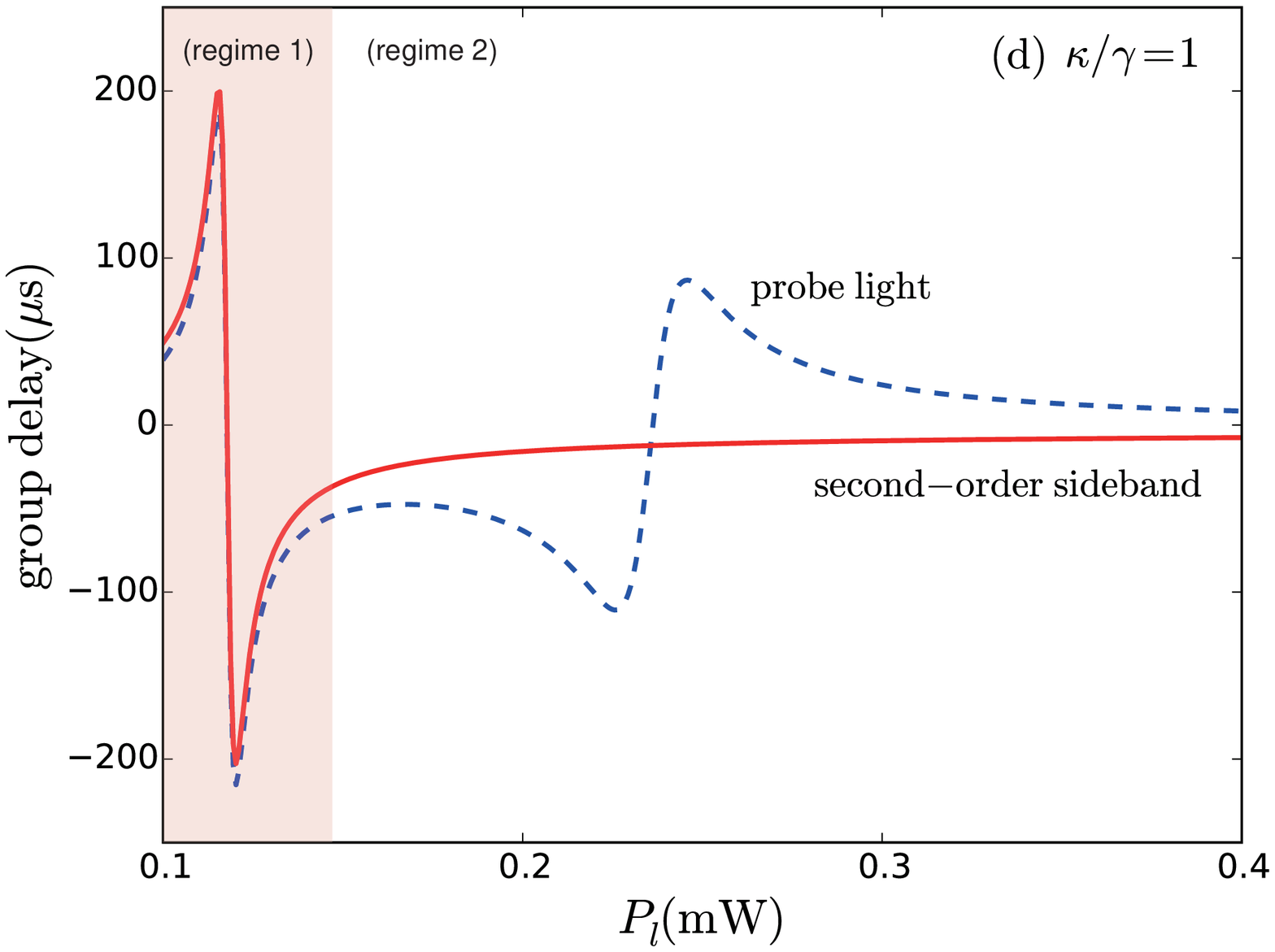}
\caption{Active or passive OMIT in a single cavity: the transmission rate of the probe $|t_{p}|^2$ (a) and the efficiency $\eta$ of the second-order sideband, with the pump power $P_{l}=933\mu$W. Here $\kappa<0$ indicates a passive cavity. The associated optical group delay at the resonance $\Delta_{p}=0$ is also plotted as a function of $P_{l}$, for a lossy cavity (c) or an active cavity (d). The regime 1 or 2 denotes the gain-dominated or the OMIT-dominated regime (see the text).}
\label{fig2}
\end{figure}

In sharp contrast to the conventional OMIT profile in a lossy cavity, an inverted-OMIT profile appears for the probe field in an active cavity (see Fig. \ref{fig2}(a)), characterizing a non-amplifying dip in the otherwise strongly amplifying region \cite{Jing2015Optomechanically,IEIT}. However, as shown in Fig. \ref{fig2}(b), the double-peak spectrum exists for $\eta$ also in the active OMIT. Nevertheless, $\eta$ is still enhanced by the gain for $\sim 60\%$ at $\Delta_{p}=0$, for $P_{l}=933\mu$W. For lower values of $P_l$, the transmission of the probe is decreased due to the weakened OMIT effect. Correspondingly, the enhanced optical absorption leads to the enhanced two-phonon up-converted process of the pump field, i.e. the increasing of $\eta$. However, as $P_l\rightarrow 0$, both of two signals are rapidly suppressed, since the whole OMIT mechanism tends to disappear \cite{Xiong2012Higher}.

The gain can also strongly affect the dispersion of the system, hence greatly altering the phases of the two signals. Fig. \ref{fig2}(c,d) shows the resulting group delays of the probe and the up-converted sideband. For low-power values, both $\tau_g$ and $\tau_g'$ are enhanced significantly due to the gain, i.e. $\sim 50$ or 100 times for their maximum values in comparison with the lossy-cavity results. In particular, a remarkable slow-to-fast light transition happens at $P_l \sim 0.12\,\mathrm{mW}$, for $\kappa/\gamma=1$, which is reminiscent of the gain-assisted fast light in an atomic gas \cite{Wang2000fastlight}. Clearly, in this weak-power regime, the OMIT effect is relatively weak and thus we observe the similar gain-dominated group delays of both two signals (see the regime 1 in Fig.\,2(d)). In contrast, for higher values of $P_l$, the more prominent OMIT effect leads to the reversed fast-to-slow light transition for the probe (see the regime 2 in Fig.\,2(d)). The relevant second-order signal has no such a transition as it tends to be relatively suppressed, as in the passive OMIT \cite{Xiong2012Higher}, leading to a tunable (negative or positive) relative delay of the two signals.

\subsection{Double-cavity case}

\begin{figure}[tbp]
\includegraphics[width=0.5\textwidth]{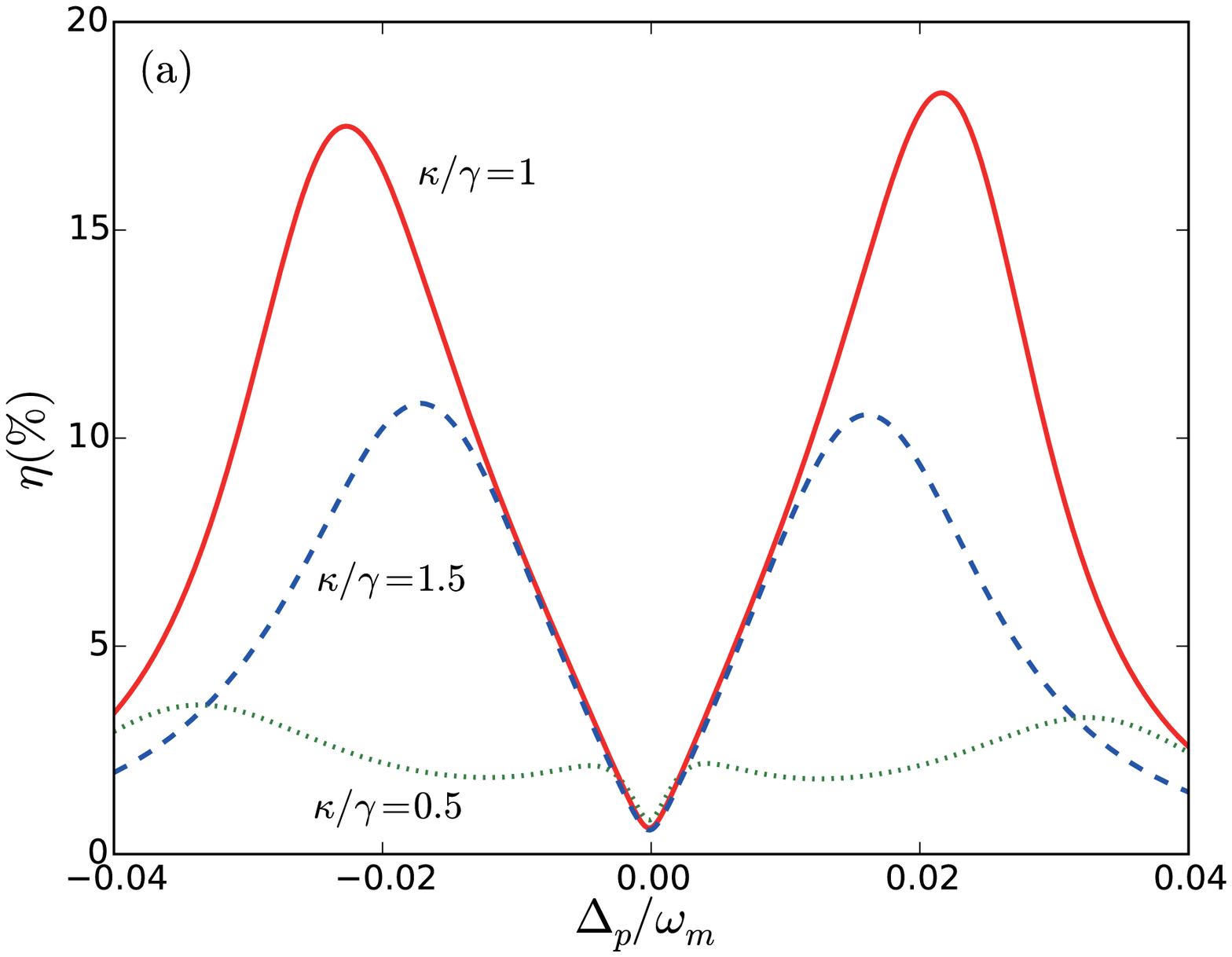}
\includegraphics[width=0.5\textwidth]{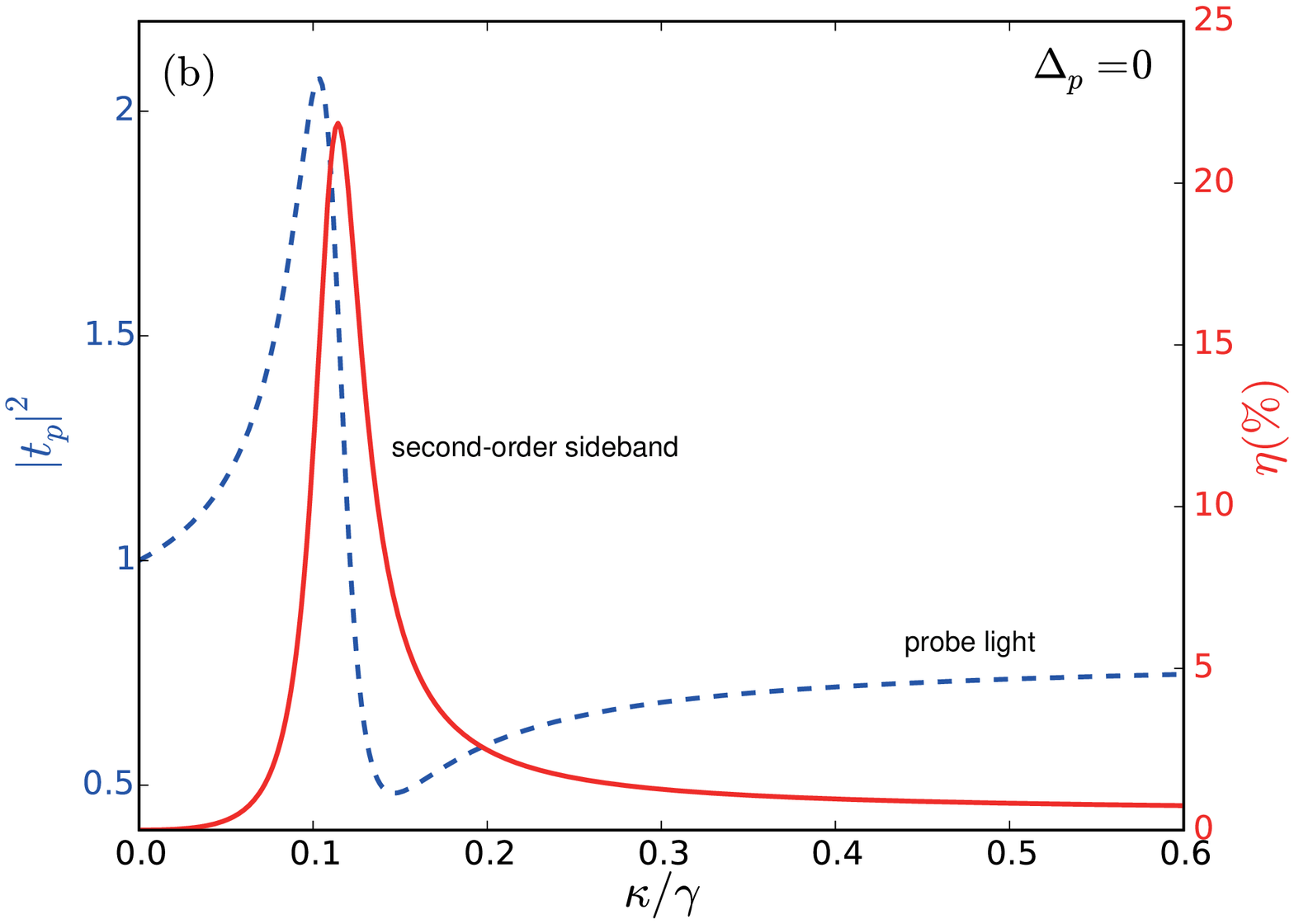}
\caption{(a) Transmission spectra of the second-order sideband in an active-passive-coupled system; (b) $|t_{p}|^2$ (blue dashed) and $\eta$ (red solid) at the resonance $\Delta_p=0$, as a function of the gain-to-loss ratio $\kappa / \gamma$. In all the figures we have taken $J/\gamma=1$ and $P_l = 933$ $\mu$W.}
\label{fig3}
\end{figure}

As shown in Figs. \ref{fig3}(a), the second-order profile is similar to the first-order case \cite{Jing2015Optomechanically}: the field amplification tends to be maximized as the gain and loss approach to the balance ($\kappa/\gamma=1$), and the reversed-gain dependence works for both processes. In fact, by taking $\omega_{m}/\omega_{c}\sim 0$, $x_{s}\sim 0$, we approximately have
\begin{equation} \label{eq:eta}
\eta \approx \left|\frac{i\hbar^2 g^4 |a_{1,s}|^2 a_{1,s}^{\ast} \kappa^4 \gamma \varepsilon_{p}}
{2m^2\Gamma_{m}^2\omega_{m}^3 (J^2-\kappa \gamma)^3 \left[m\omega_{m}\Gamma_{m}(J^2-\kappa \gamma) + i\hbar g^2|a_{1,s}|^2(J^2-\kappa^2)\right]}\right|,
\end{equation}
i.e., $\eta$ is most preferable in the vicinity of the exceptional point (EP) at the gain-loss balance \cite{Peng2014Parity,Jing2015Optomechanically}, i.e. $\kappa/\gamma=1$, with $J/\gamma=1$. Clearly, for $\kappa/\gamma<1$, the loss in the passive cavity can be compensated by the gain, leading to the sideband amplification of the transmission spectrum; after passing by the EP, the optical supermodes become spontaneously localized in either the amplifying or the lossy cavity \cite{Peng2014Parity}, thus the loss of the passive cavity is increased again. This, of course, does not means that $\eta$ is maximized for $\kappa/\gamma=1$ at all detuning values \cite{El-Ganainy}. Still, at $\Delta_p=0$, we have $\sim 3$ times enhancement of $\eta$ by tuning $\kappa/\gamma$, in comparison with even the single-active-cavity situation (see Fig.3(b)), indicating the advantage of using a compound system.

\begin{figure}[tbp]
\centering
\includegraphics[width=0.49\textwidth]{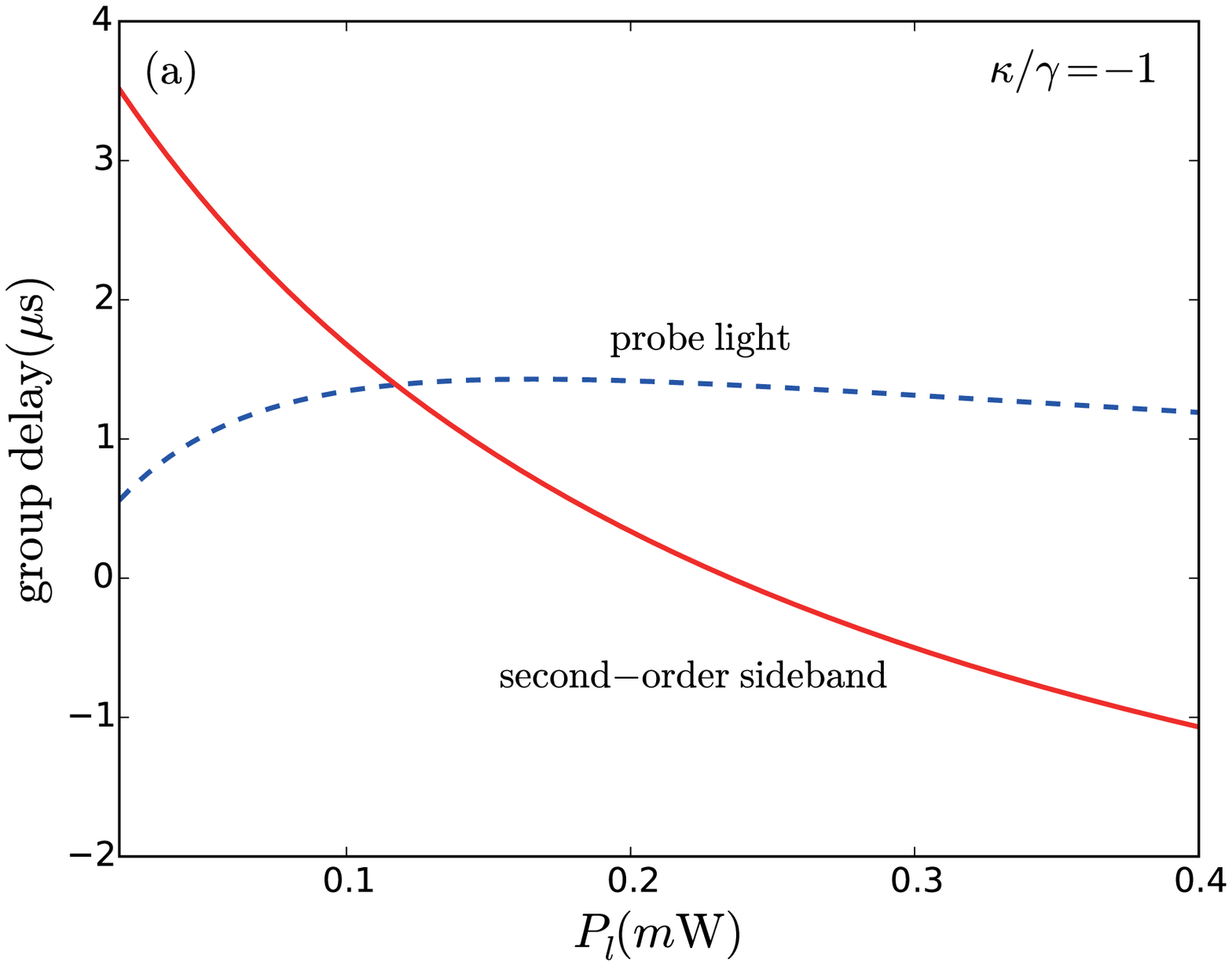}
\includegraphics[width=0.49\textwidth]{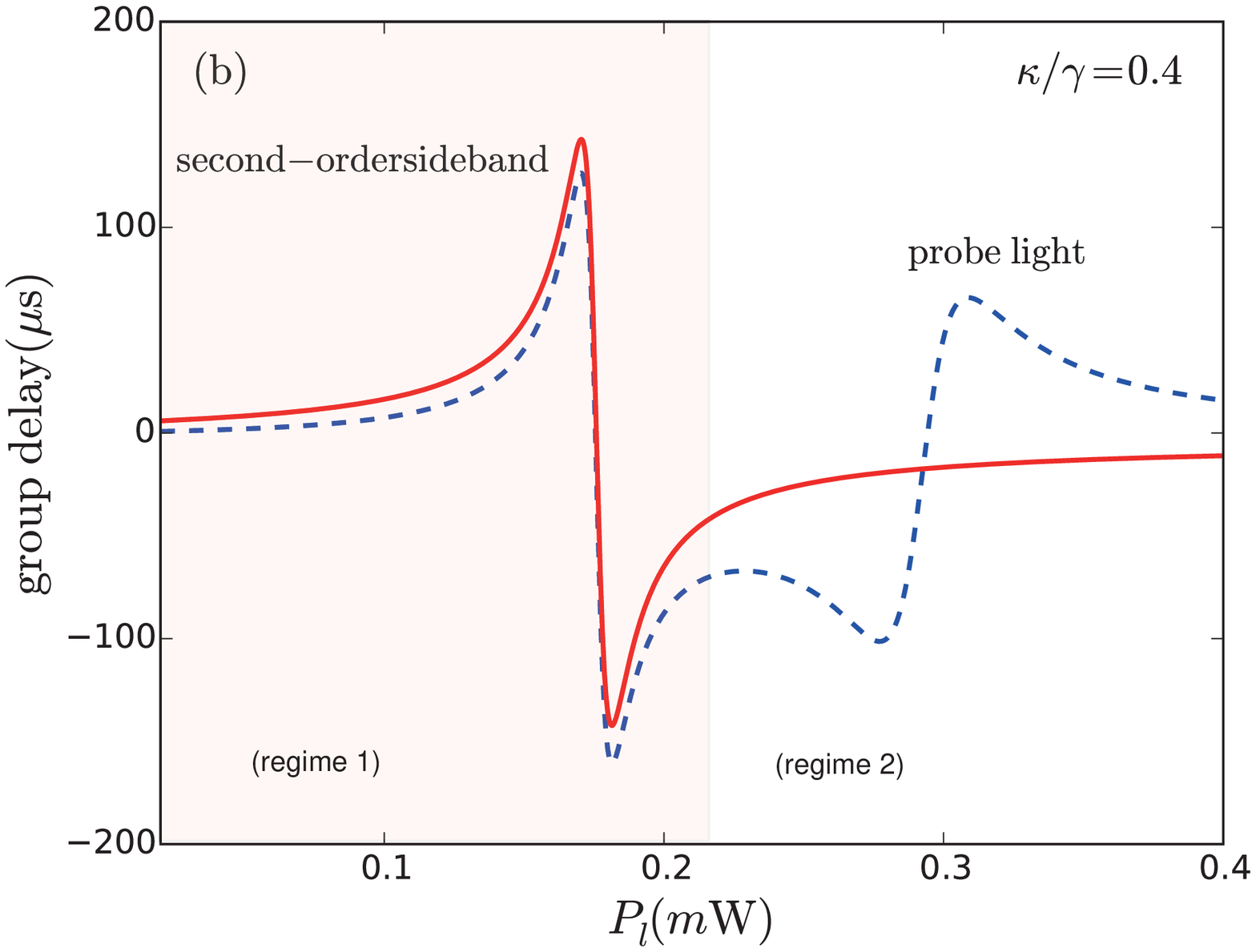}  \\
\includegraphics[width=0.49\textwidth]{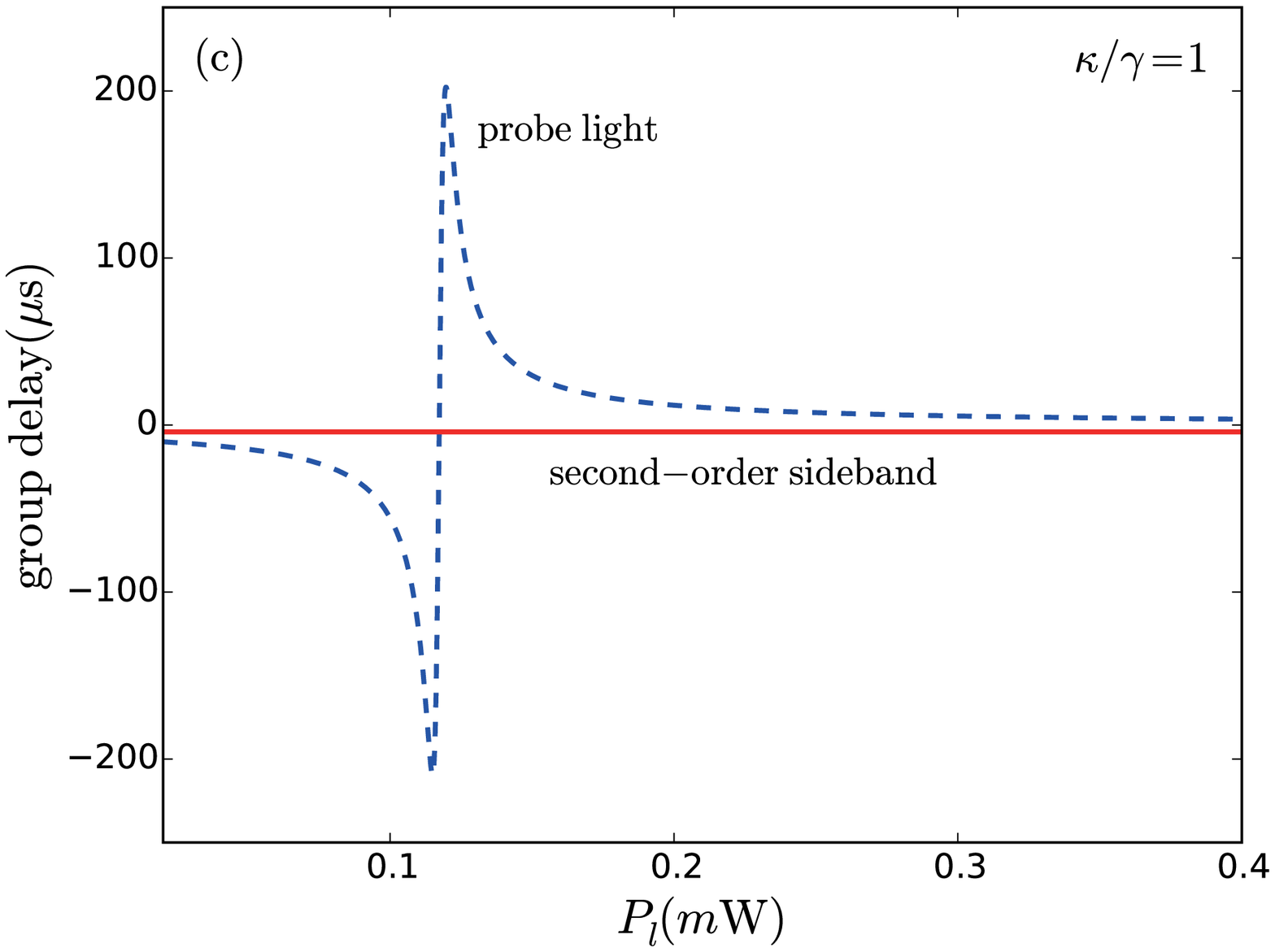}
\includegraphics[width=0.49\textwidth]{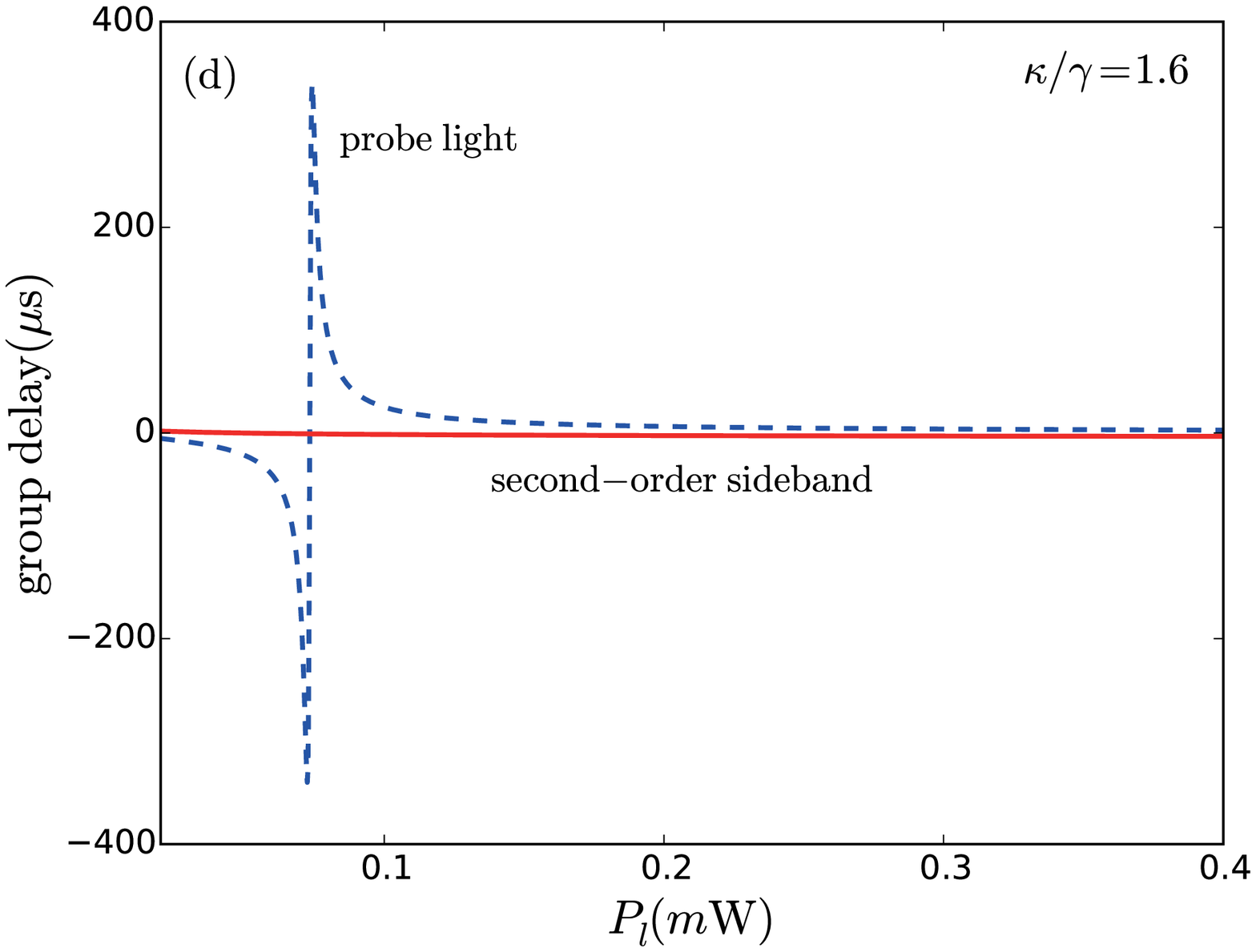}
\caption{Optical group delay at the resonance $\Delta_{p}=0$, as a function of pump power for the passive-passive system ($\kappa/\gamma=-1$) (a) and the active-passive system with $\kappa/\gamma=0.4$ (b), 1.0 (c), or 1.6 (d).}
\label{fig4}
\end{figure}

The associated optical group delay is shown in Fig. \ref{fig4}. For the coupled passive resonators ($\kappa/\gamma=-1$), $\tau_g$ and $\tau^{\prime}_g$ changes within the scope of only several microseconds, as in the single-passive-cavity case (see Fig.\,4(a)). For $\kappa/\gamma<1$, due to the loss compensation by gain, similar results as in the single-active-cavity case can be observed, i.e. the gain-dominated low-power regime, with similarly giant enhancement of both $\tau_g$ and $\tau^{\prime}_g$ (around the slow-to-fast light transition, see Fig.\,4(b)), which is then followed by the OMIT-dominated regime for higher values of $P_l$. In contrast, for $\kappa/\gamma\geq 1$, as shown in Ref.\,\cite{Jing2015Optomechanically}, the regime 1 as shown in Fig.\,4(b) now disappears, since the slow-to-fast light transition in the low-power regime is reversed to the fast-to-slow light transition, a feature characterizing the unconventional PT-broken regime (see Fig.\,4(c-d)) \cite{Jing2015Optomechanically}. Accompanying with this reversal, the group delay of the second-order signal tends to be rapidly suppressed in the PT-broken regime. This feature does not exist at all in any single cavity case, in which the relative group delay of the probe and the second-order signal becomes significant only for higher values of the pump power (see e.g. Fig.\,2(d)). In practice, this may provide highly sensitive low-power diagnoses of different phases of the OMIT systems.

\section{Conclusion} \label{sec:conclusion}

In summary, we have studied unconventional effects in nonlinear OMIT with gain and loss, including the inverted-OMIT spectrum for the second-order sideband and the tunable optical delay. We note that recently, for a single cavity, the gain-enhanced COM cooling was revealed \cite{Geli}, but the OMIT effects, i.e. the optical transparency or the group delay, was not yet studied. We also note that in COM, coupled resonators can possess several advantages in comparison with the single cavity \cite{phononlaser2,LvXY2015}. The previous work on active OMIT was limited to the linear process in coupled resonators \cite{LvXY2015}; hence our work here enables comparisons not only of the single-resonator and coupled-resonator cases, but also of the linear and nonlinear OMIT processes.

In the practical aspect, the active COM device can generate a frequency-shifted sideband, which is similar to an acousto-optical modulator, with also an additional ability to separate it with the input signal in timing sequence. Further studies of this aspect is, however, beyond the purpose of our work here. In the fundamental aspect, our study indicates the novel possibility to actively enhance other important nonlinear COM effects, such as photon blockade \cite{Rabl1,Rabl2}, phonon squeezing \cite{Wollman2015Quantum}, and photon-phonon entanglement. In the future, we also plan to study nonlinear OMIT in a system with multiple phonon modes \cite{Fan2015Cascaded} or intrinsic two-level-system defects \cite{spin}, especially the possibility of enhancing the sensitivity of detecting the inter-mode mechanical coupling or the phonon-defect coupling, by virtue of the actively amplified OMIT spectrum \cite{Liu,Xuereb1,Xuereb2}.

\section*{Acknowledgements}

We thank Xin-You L\"u for helpful discussions. This work is supported partially by the CAS 100-Talent program and the National Natural Science Foundation of China under Grant numbers 11274098, 11474087, and 11422437.

\bibliographystyle{iopart-num}

\end{document}